\documentclass[seceq]{ptptex}

\usepackage{graphicx}
\def\gsi{\raise0.3ex\hbox{$>$\kern-0.75em\raise-1.1ex\hbox{$\sim$}}}
\newcommand{\gsim}{\mathop{\gsi}}

\title{
{\normalsize\vspace*{-1.9cm}\\\hfill\mbox{ITP-Budapest 606}\\
\hfill\mbox{WUB 04-02}\\\vspace*{0.6cm}}
The QCD equation of state at finite $T/\mu$ on the lattice
}

\author{
F.~{\sc Csikor},$^{1,}$ 
G.~I.~{\sc  Egri},$^{1}$ 
Z.~{\sc Fodor},$^{1,2}$ 
S.~D.~{\sc Katz},$^{2}$ \footnote{presented by S.~D.~Katz} \footnote{On leave from Inst. Theor. Phys., E\"otv\"os Univ.}
K.~K.~{\sc Szab\'o}$^{1}$ and 
A.~I.~{\sc T\'oth}$^{1}$ 
}

\inst{
$^{1}$ Institute for Theoretical Physics, E\"otv\"os
University, P\'azm\'any P. 1/A, H-1117 Budapest, Hungary.
\\
$^{2}$ Deapartment of Physics, University of Wuppertal, 
D-42097 Wuppertal, Germany
}

\abst{
We present $N_t=4$ lattice results for the equation of
state of ~$2+1$ flavour staggered, dynamical QCD at finite temperature
and chemical potential.  We use the overlap improving multi-parameter
reweighting technique to extend the equation of state for
non-vanishing chemical potentials.  The results are obtained along the
line of constant physics. Our physical parameters extend in
temperature and baryon  chemical potential upto $\approx 500-600$~MeV.
}

\begin{document}

\maketitle

\section{Introduction}
\label{sec:intro}

QCD at finite $T$ and/or $\mu$ is of special importance since it can
be used to describe the early universe, neutron stars and also 
heavy ion collisions. Present and future heavy ion collisions are
carried out at CERN, in Brookhaven and at GSI to detect and study
experimentally the QGP phase, i.e. QCD at large temperature and
moderate chemical potentials \cite{Wilczek:1999ym,Kogut:2002kk}.
It is very important to 
understand the theoretical grounds of the underlying physics. First
principle answers can be gained from lattice QCD calculations e.g. to
obtain the equation of state (EoS). 

The experiments are carried out at $\mu\neq 0$ but unfortunately 
until recently the lattice results were limited to 
$\mu=0$. Though lattice QCD can be easily formulated at non-vanishing
chemical potentials \cite{Hasenfratz:1983ba,Kogut:1983ia}
we cannot use Monte-Carlo
simulations at $\mu\neq 0$ as the determinant of the  Euclidean Dirac
operator and so the functional measure becomes complex.

Recently two of us proposed a new technique
\cite{Fodor:2001au}, the so-called overlap
improving multi-parameter reweighting method to study lattice QCD at
finite $\mu$ . This procedure proved to be good enough to give the
phase boundary on the $T$-$\mu$ plane for four flavours 
\cite{Fodor:2001au}, for 2+1 flavours \cite{Fodor:2002pe} and the 
equation of state \cite{Fodor:rovid,Fodor:long}.
Essentially the same technique was used succesfully by other studies 
\cite{Allton:2002zi,Karsch:eos,Choe:2002mt}; however,
instead of evaluating the fermionic determinant exactly it can be
approximated by its Taylor series with respect to $\mu$.
Other approaches, like
simulations at imaginary chemical potential and analytic continuation
led to results that are in good agreement with those of our method
\cite{deForcrand:2002ci,deForcrand:2003hx,D'Elia:2002gd}. 

In this paper we determine the EoS on the line of constant physics
(LCP). An LCP can be defined by a fixed ratio of the strange quark
mass ($m_s$) and the light quark masses ($m_ud$) to the $\mu=0$
transition temperature ($T_c$). Our parameter choice approximately 
corresponds to the physical strange quark mass. However, the ratio of
the pion mass ($m_{\pi}$) and the rho mass ($m_{\rho}$)
is around 0.5-0.75, which is roughly 3 times larger than its physical
value. In our lattice analysis we use $2+1$ flavour QCD with dynamical
staggered quarks. The determination of the EoS at finite chemical
potential needs several observables at non-vanishing $\mu$-s. These
are produced by the use of the multi-parameter reweighting method. 
We employ the integral method to calculate the pressure
\cite{Engels:1990vr}.

The paper is organized as follows. In Section 2 we summarize the
lattice parameters and the technique by which
the lines of constant physics can be determined. Section 3 presents the
equation of state at vanishing chemical potential. Sections 4 deals
with the question how to reweight into the region of $\mu\neq 0$ and 
how to estimate the error of the reweighted quantities. 
In Section 5 we give the equation of state for non-vanishing chemical
potential and temperature. Those who are not interested in the details
of the lattice techniques should simply omit Sections 2--4 and jump to
Section 5, or refer to \cite{Fodor:rovid}.
Finally, Section 6 contains a summary and the conclusions.
The details of this work can be found in~\cite{Fodor:long}. 

\section{Lattice parameters and the line of constant physics}
\label{sec:lat_param}

In this paper we use $2+1$ flavour dynamical QCD with unimproved
staggered action. Simulations are done for the equation of state along two
different lines of constant physics and at 14 different temperatures.
The temperature range spans up to $3 T_c$.  In physical units our 
parameters correspond to pion to rho mass ratio of $m_\pi/m_\rho 
\approx 0.5 - 0.75$ and lattice spacings of $a \approx 0.12-0.35$~fm. 

The finite temperature contributions to the EoS are obtained on
$4\cdot 8^3$, $4\cdot 10^3$ and $4\cdot 12^3$ lattices, which can be
used to extrapolate into the thermodynamical limit (we usually call
them hot lattices).  On these lattices we determine not only the usual
observables (plaquette, Polyakov line, chiral condensates) but also
the determinant of the fermion matrix and the baryon density ($n_B$)
at finite $\mu$. $10000-20000$ trajectories are simulated at each bare
parameter set. Plaquettes, Polyakov lines and the chiral condensates
are measured at each trajectories whereas the CPU demanding
determinants and related quantities are evaulated at every 30
trajectories. For our parameters the CPU time used for the production
of configurations is of the same order of magnitude as the CPU time
used for calculating the determinants.

Since we usually move along the line of constant physics by changing
the lattice spacing $a$ and keeping the masses fixed we will
explicitely write out the lattice spacing $a$ in our formulas.  In
this paper we study lattices with isotropic couplings.  
We write $\mu_B$ for the baryonic chemical potential,
whereas for the quark chemical potential ($u,d$ quarks) we use the
notation $\mu$.  Similarly, the baryon density is denoted by $n_B $
and the light quark density by $n$.

In the remaining part of this 
section we discuss the role of LCP when determining the EoS in
pure gauge theory and in dynamical QCD. After that we determine the
lines of constant physics, along which our simulations are done.

In order to detetermine the temperature $T=1/[N_ta(\beta)]$ of the
pure gauge theory, we have to compute the lattice spacing ($a$) as a
function of the gauge coupling ($\beta$). 
In the $d$ dimensional space of the bare
parameters one defines $d$ appropriately chosen quantities. The LCP is
given by $d-1$ constraints and it is parametrized by a non-constrained
combination of the above quantities.  For the $2+1$ flavour staggered
action we have three bare parameters ($\beta$, and two masses,
$m_{ud},m_s$). Thus, we need two constraints. There are several
possibilities for these constraints and consequently there are many
ways to define an LCP.  A convenient choice for two of the three
quantities can be the bare quark masses ($m_{ud}$ and $m_s$). A more
physical possibility is to use the pion and kaon masses
($m_\pi,m_K$). 

In our analysis first we use the bare quark masses ($m_{ud}$ and
$m_s$) and the transition temperature to define an LCP. In this paper
we use two\footnote{As we will see later, two LCP's are needed for the
determination of the EoS at finite chemical potential.} 
LCP's (LCP$_1$ and LCP$_2$).  The conditions
\begin{eqnarray}\label{constraint}
\begin{array}{c}
m_{ud}=0.48T_c =0.48/(N_ta)  \ \ \ \ {\rm and}\ \ \ \ \  
m_s=2.08\cdot m_{ud}\\
m_{ud}=0.384T_c=0.384/(N_ta) \ \ \ \ {\rm and}\ \ \ \ \  
m_s=2.08\cdot m_{ud}
\end{array}
\end{eqnarray}
are taken as the constraints for LCP$_1$ and LCP$_2$,
respectively. For both LCP's we determined four
different transition couplings ($\beta_c$) by susceptibility peaks on
$N_t=1/(T_ca)=4$, 6, 8 and 10 lattices with spatial extensions
$N_s\gsim 3N_t/2$ and quark masses given by eq.~(\ref{constraint}).
The quark masses or the transition gauge couplings can be used to
parametrize the LCP's.

\begin{figure}[t]
\centerline{\includegraphics[width=8cm]{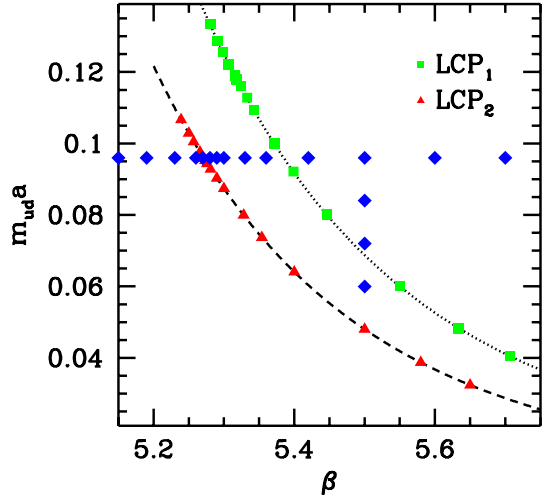}} 
\vspace{-2mm}
\caption{The lines of constant physics (LCP$_1$ and
LCP$_2$) on the $\beta$ vs. $m_{ud}a$ plane.  The strange quark mass
is given by $m_s=2.08 m_{ud}$ for both LCP's. The simulation points are
shown by squares/triangles and connected by dashed/dotted lines for
LCP$_1$/LCP$_2$, respectively. The diamonds
along a horizontal line represent the simulation points in the
``non-LCP'' approach. Additional 4 diamonds in the vertical direction
show the simulation points used to test the path independence of the
integral method.
\label{LCP}}
\end{figure}

By the finite temperature technique, described above, only a few
points of the LCP's can be obtained. To interpolate $\beta(a)$ between
these points (and extrapolate slightly away from them) we use the
renormalization group inspired ansatz proposed by Allton
\cite{Allton:1996kr}. A particularly
illustrative parametrization is obtained by inverting
eq.~(\ref{constraint}) and using $N_t$ as a continuous parameter. 
Figure \ref{LCP} shows LCP$_1$ and LCP$_2$ with our
simulation points.  The simulation points in the
``non-LCP'' approach -- often used in the literature --  are also shown.  Note that even though the
determination of the LCP$_1$ and LCP$_2$ are done on finite
temperature lattices, the obtained bare parameters are used in the
rest of the paper for $T=0$ and $T \neq 0$ simulations.

\section{Equation of state along the lines of constant physics (LCP) at $\mu=0$}
\label{sec_mu0}

In previous studies of the EoS with staggered quark actions, the
pressure and the energy density were determined as functions of the
temperature for fixed value of the bare quark mass $m_qa$   in the 
lattice action \cite{Bernard:1997cs,Engels:1997ag,Karsch:2000ps}. In
these studies a fixed $N_t$ was used (e.g. $N_t = 4$ or 6) at different
temperatures.  Since $T=1/(N_t a)$ the temperature is set by the
lattice spacing which changes with $\beta$.  This convenient, fixed
bare $m_qa$ choice leads to a system which has larger and larger
physical quark masses at decreasing lattice spacings (thus, at
increasing temperatures). Increasing physical quark masses with
increasing temperatures could result in systematic errors of the EoS.

Clearly, instead of this sort of analysis (in the rest of the paper we
refer to it as ``non-LCP approach'') one intends to study the
temperature dependence of a system with fixed physical observables,
therefore on an LCP.

In our analysis we use full QCD with staggered quarks along the LCP
and compare these results with those of the ``non-LCP approach''.

Now, we briefly review the basic formulas and emphasize the issues 
related to the EoS determination along an LCP.

The energy density and pressure are defined in terms of the
free-energy density ($f$):
\begin{eqnarray}
\epsilon (T)=f-T \frac{\partial{f}}{\partial {T}},\ \ \ \ \ \ \ \ && p(T) = -f.
\end{eqnarray} 
Expressing the free energy in terms of the partition function 
($f=-T/V \log Z =$ \\ $ -T \partial (\log Z)/\partial V$) we have:
\begin{eqnarray} \label{eps_p}
\epsilon (T)=\frac{T^2}{V}\frac{\partial{\log Z}}{\partial {T}}, \ \ \
\ \ \ \ \ && 
p(T) = T\frac{\partial (\log Z)}{\partial V}.
\end{eqnarray} 

The temperature and volume are connected to this lattice spacing by
\begin{eqnarray} \label{TV}
T=\frac{1}{a N_t}, && V=a^3 N_s^3.
\end{eqnarray}

Inspecting eqs.(\ref{eps_p}, \ref{TV}) we see that 
$(\epsilon-3p)/T^4$ is directly proportional to the total
derivative of $\log Z$ with respect to the lattice spacing:
\begin{equation}\label{interaction}
\frac{\epsilon-3p}{T^4}=-\frac{N_t^3}{N_s^3}a \frac{d (\log Z)}{d a}.
\end{equation}
Here, the derivative with respect to $a$ is defined along the LCP,
which means that only the lattice spacing changes and the physics (in
our case $m_q/T_c$) remains the same. We can write:
\begin{equation}
\frac{d}{da}=\frac{\partial \beta}{\partial a}\frac{\partial}{\partial \beta} +
\sum_q \frac{\partial (m_q a)}{\partial a}\frac{\partial}{\partial (m_q a)}.
\end{equation}
Since the LCP is defined by $m_q/T_c={\rm const.}$, the
partial derivative $\partial (m_qa)/\partial a$ becomes simply $m_q$.
The derivatives of $\log Z$ with respect to $\beta$ and $m_q$ are the
plaquette and $\bar{\Psi}\Psi_q$ averages multiplied by the lattice
volume. We get:
\begin{equation}\label{interaction-measure}
\frac{\epsilon-3p}{T^4}=-N_t^4 a 
\left(\overline{{\rm Pl}} \left. \frac{\partial \beta}{\partial a}\right|_{\rm  LCP}+
\sum_q \overline{\bar{\Psi}\Psi}_{q} m_{q} \right).
\end{equation}

The pressure is usually determined
by the integral method \cite{Engels:1990vr}.  The pressure is simply
proportional to $\log Z$, however it cannot be measured directly. One
can determine its partial derivatives with respect to the bare
parameters. Thus, we can write:
\begin{equation}\label{integral}
\frac{p}{T^4}=\left[-\frac{N_t^3}{N_s^3}\int^{(\beta,m_q a)}_{(\beta_0,m_{q0} a)}
d (\beta,m_q a) 
\left(\begin{array}{c}
{\partial \log Z}/{\partial \beta} \\
{\partial \log Z}/{\partial (m_q a)}
\end{array} \right )\right]- \frac{p_0}{T^4}.
\end{equation}
Since the integrand is the gradient of $\log Z$, the result is by
definition independent of the integration path (we explicitely checked this
path independence). For the substracted vacuum term we used the zero 
temperature pressure, i.e. the same integral on $N_{t0}= N_s$ lattices.  
The lower limits of the integrations (indicated by $\beta_0$ and $m_{q0}$) 
were set sufficiently below the transition point. By this choice the 
pressure becomes independent of the starting point (in other words it 
vanishes at vanishing temperature).  In the case of $2+1$ staggered QCD eq. 
(\ref{integral}) can be rewritten appropriately and the pressure is given by
\begin{equation}\label{pmu0}
\frac{p}{T^4}=
-N_t^4\int^{(\beta,m_q a)}_{(\beta_0,m_{q0} a)}
d (\beta,m_{ud} a,m_s a) 
\left(\begin{array}{c}
\langle{\rm Pl}\rangle \\
\langle\bar{\Psi}\Psi_{ud}\rangle \\
\langle\bar{\Psi}\Psi_{s}\rangle
\end{array} \right),
\end{equation}
where we use the following notation for subtracting the vacuum term:
\begin{equation}
\langle  {\cal O}(\beta,m) \rangle= 
{\overline {{\cal {O}}}(\beta,m)}_{T\neq 0}-
{\overline {{\cal O}}(\beta,m)}_{T=0}.
\end{equation}

\begin{figure}[t]
\centerline{\includegraphics[width=14cm]{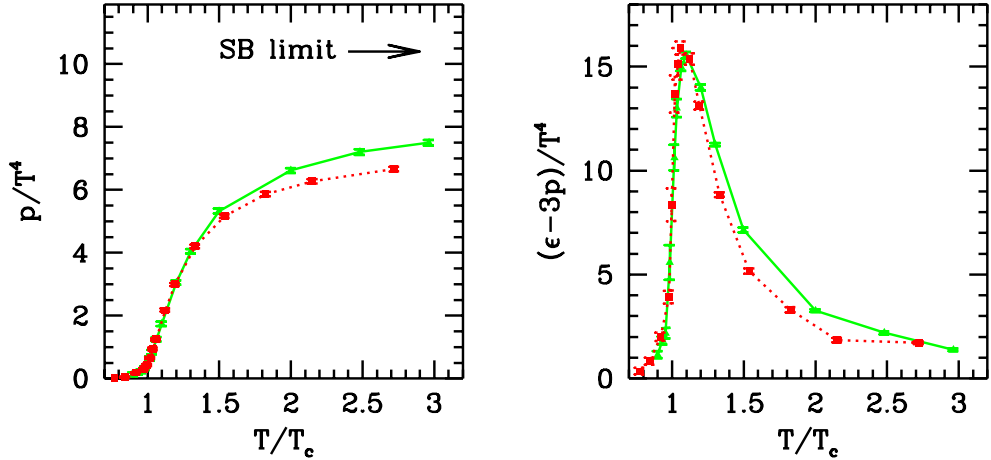}} 
\vspace{-2mm}
\caption{The equation of state at $\mu$=0. (a) The left
panel shows the pressure $p$, as a function of the temperature.  All
quantities are normalised by $T^4$. In order to lead the eye the
solid line connects the data points obtained along the LCP$_2$,
whereas the dashed line connects the data points obtained in the
``non-LCP approach'' (see text). The Stefan-Boltzmann limit is also shown by 
an arrow for $N_t$=4 lattices. The EoS along the LCP and the ``non-LCP 
approach'' differ from each other at high T. (b) The same for $\epsilon$-3p.
\label{eosmu0}}
\end{figure}

Figure \ref{eosmu0} shows the EoS at vanishing chemical potential on 
$N_t =4$ lattices for LCP$_2$ and for the non-LCP approach. The pressure 
and $\epsilon-3p$ are presented as a function of the temperature. The 
parameters of LCP$_2$ and those of the non-LCP approaches coincide at 
$T = T_c$. The Stefan-Boltzmann limit valid for $N_t = 4$ lattices is 
also shown.

It can be seen that the EoS along the LCP and the ``non-LCP approach''
differ from each other at high temperatures. This is an obvious
consequence of the fact that in the ``non-LCP approach'' the bare
quark mass increases linearly with the temperature. The ``LCP'' pressure 
is much closer to the SB limit. 

\section{Reliability of reweighting and the best reweighting lines}
\label{sec_rely}

The aim of this section is to study the reliability of the multi-parameter 
reweighting (for another study of reweighting see~\cite{Ejiri:2004yw}). We also determine its region of validity by a suitably estimated 
error. To start with let us briefly review the  multi-parameter reweighting. 

As proposed in \cite{Fodor:2001au} one can identically rewrite the partition 
function in the form:
\begin{eqnarray}\label{multi-parameter}
Z(m,\mu,\beta) =
\int {\cal D}U\exp[-S_{bos}(\beta_0,U)]\det M(m_0,\mu=0,U)\\
\left\{\exp[-S_{bos}(\beta,U)+S_{bos}(\beta_0,U)]
\frac{\det M(m,\mu,U)}{\det M(m_0,\mu=0,U)}\right\},\nonumber
\end{eqnarray}
where $U$ denotes the gauge field links and $M$ is the fermion matrix 
\footnote{For $n_f\neq 4$ staggered dynamical QCD one simply takes 
fractional powers of the fermion determinant.}. The chemical potential 
$\mu$ is included as $\exp(a\mu)$ and $\exp(-a\mu)$ multiplicative 
factors of the forward and backward timelike links, respectively. In 
this approach we treat the terms in the curly bracket as an observable 
-- which is measured on each independent configuration, and can be 
interpreted as a weight -- and the rest as the measure.  Thus the  simulation 
can be performed at $\mu=0$ and at some $\beta_0$ and $m_0$ values 
(Monte-Carlo parameter set). By using the  reweighting 
formula~(\ref{multi-parameter}) 
one obtains the partition function at another set of parameters, 
thus at $\mu\neq 0$, $\beta\neq\beta_0$ or even at $m\neq m_0$ 
(target parameter set). \footnote{Note the appearent similarity 
of the present method with those of 
\cite{Ferrenberg:1988yz,Ferrenberg:1989ui,Barbour:1998ej,Csikor:1998eu,
Aoki:1999fi}.}

Expectation values of observables can be determined by the above technique. 
In terms of the weights (i.e. the expression in the curly bracket of 
eq.~(\ref{multi-parameter})) the averages can be determined as:
\begin{equation} \label{multi-parameter2}
{\overline {\cal O}}(\beta,\mu,m)=\frac{\sum \{w(\beta,\mu,m,U)\}
{\cal O}(\beta,\mu,m,U)}{\sum
\{w(\beta,\mu,m,U)\}}.
\end{equation}

Now we present an error estimate of the reweighting procedure which also 
shows how far we can reweight in the target parameter space \footnote{
For other techniques to estimate the errors of the reweighting method
see e.g. \cite{Ferrenberg:1995,Newman}}. 

The steps of the new procedure are as follows. First, we assign to each 
configuration of our initial sample the weight $w$ valid at the chosen target 
parameter set i.e.   
\begin{equation}
w=\exp\left\{\Delta\beta\cdot V\cdot(Pl)+\frac{n_f}{4}[\ln\det M(\mu)-\ln\det M(\mu=0)]\right\}
\label{eq:newweight}
\end{equation} 
where $V$ is the lattice volume. Next, we carry out 
Metropolis-like accept/reject steps with the series of these new weights. 
This procedure generates a new, somewhat smaller, sample. The configurations 
of this new sample  are taken with a unit weight to calculate expectation 
values, variances and integrated autocorrelation times \cite{Ferrenberg:1995} 
(the latter grows at every rejection due to the repetition of certain 
configurations). The expectation values are taken from 
eq.~(\ref{multi-parameter2}), whereas the errors are  
estimated from the new procedure.

We still have to clarify two important points: thermalization and the
possible lack of relevant configurations.
We solved the first problem  by defining a thermalization 
segment at the beginning of every newly generated sample which we cut 
off from the sample before calculating the expectation values and the 
errors. An obvious assumption is to claim  that the already thermalized 
sample containing valuable information starts with the first different 
configuration right after the one with the largest weight. This ensures that  
if there is only one configuration in the initial 
sample which ``counts'' at the target parameters then this information 
will not be lost. The second problem cannot be solved perfectly. This 
problem occurs e.g. when a phase transition is very strong. 

\begin{figure}[t]
\centerline{\includegraphics[width=7cm]{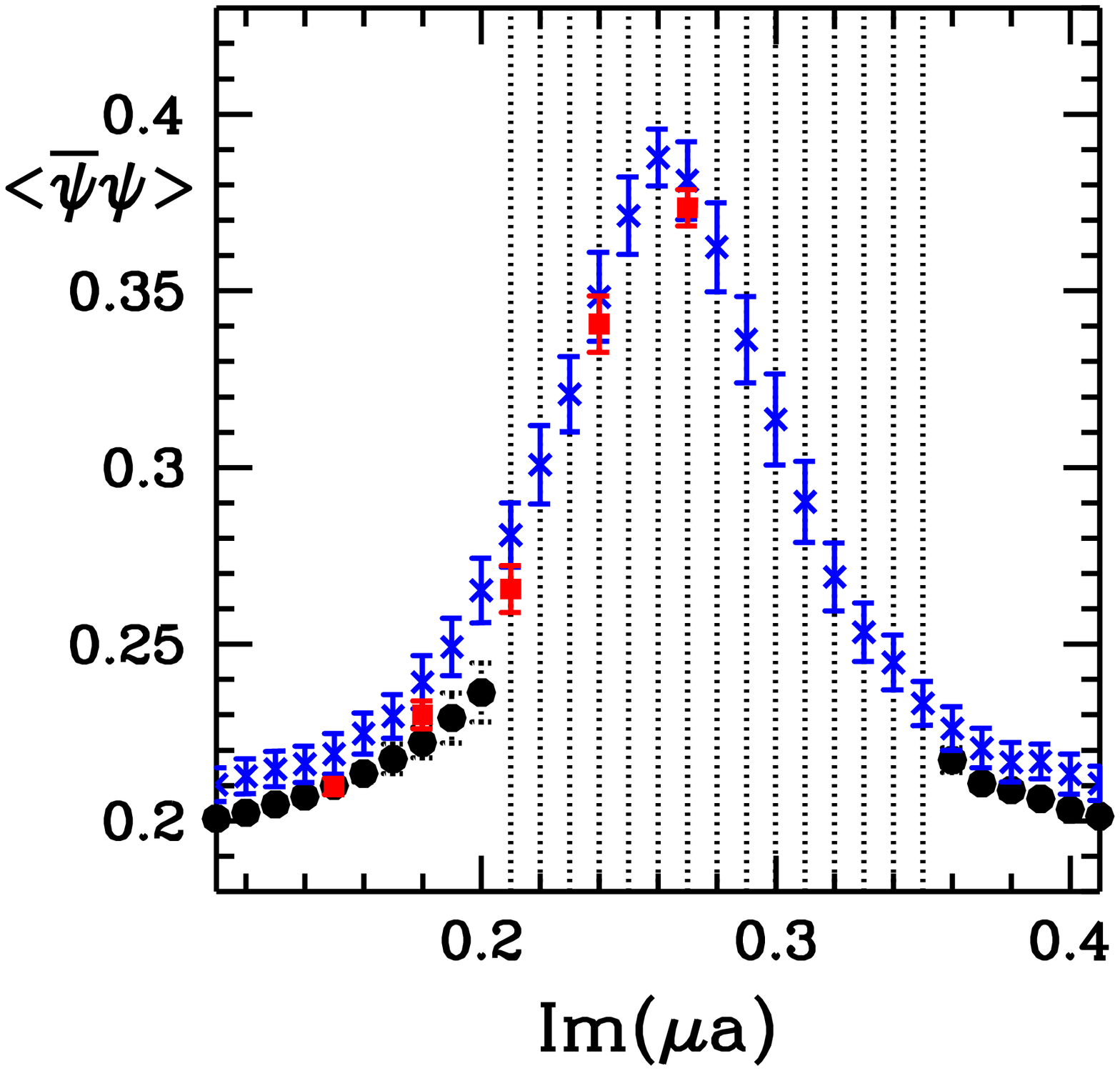} \hspace{5mm}
            \includegraphics[width=7cm]{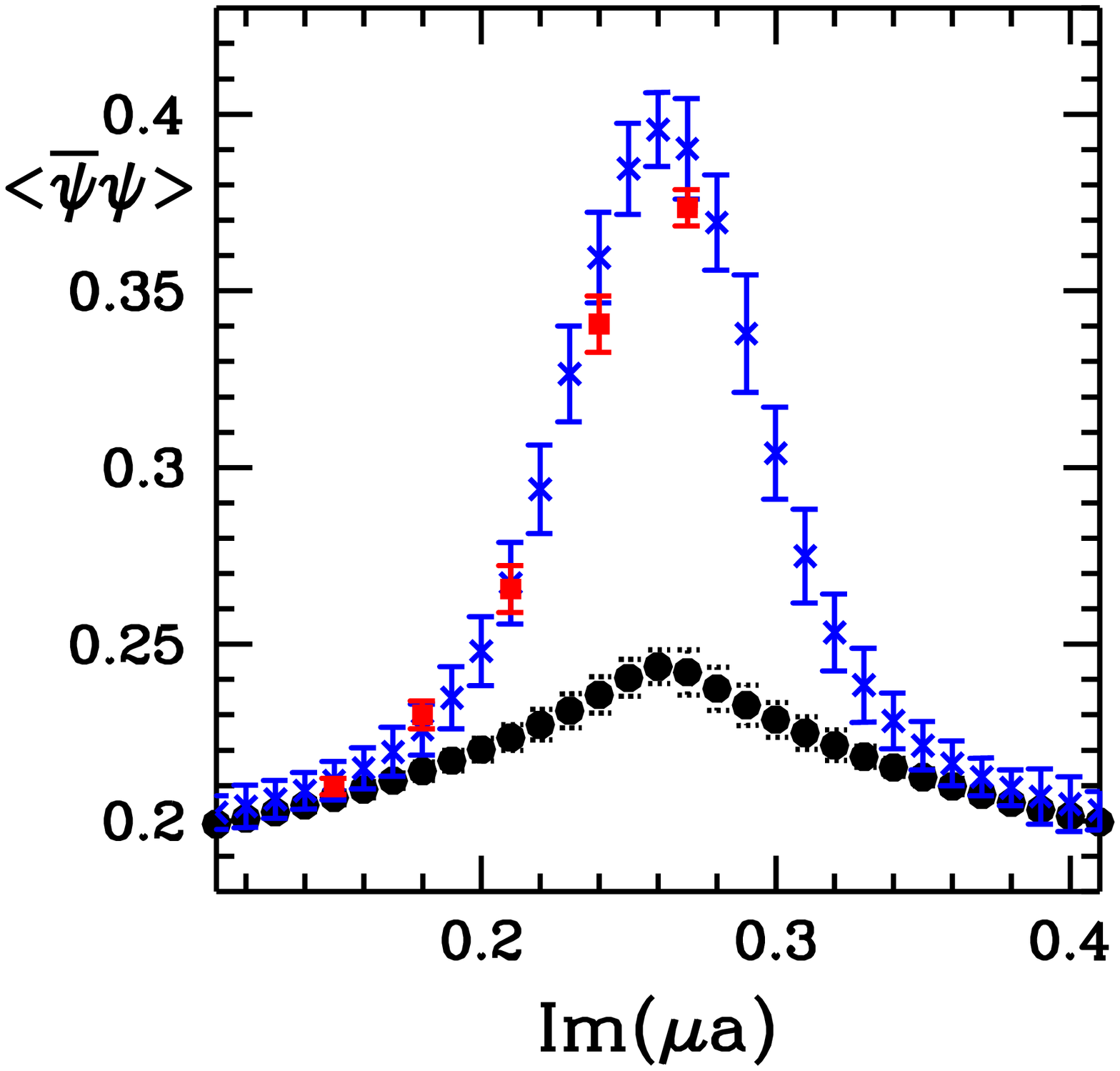}} 
\vspace{-2mm}
\caption{
(a) The left panel shows a well functioning error estimate. 
The squares correspond 
to direct simulations (out of $60000$ configurations), the crosses 
denote the results of the multi-parameter reweighting method 
(out of $\approx 2500$ 
independent configurations) while the circles are the points of the 
Glasgow-type reweighting (out of $\approx 7000$ independent 
configurations).  (b) In the right panel the meaning of the 
symbols are unchanged but  the sample sizes for the reweighting 
techniques are decreased to 1200 independent configurations. Noticeably, 
in case of the Glasgow-type reweighting the fact that the sample does 
not contain any configuration from the target phase causes systematic 
errors in the expectation values and mainly in their uncertainties. 
This is because in case of strong phase transitions -- like in our case 
-- it is far too difficult to define a reliable thermalization stage 
by  virtue of the weights. By increasing the sample size helps in this 
situation leading to the results of the left panel.
\label{fig:metro1}}
\end{figure}

To illustrate the new technique let us take a look at the $n_f=4$ 
flavour case at $m_qa=0.05$ bare quark mass 
on $4\cdot 6^3$ size lattice at imaginary chemical potential. 
Note that for purely imaginary chemical potentials direct simulation 
is possible, therefore it is possible to check the validity of any 
possible error estimation method.
We carried 
out simulations\footnote{This parameter set is identical to the one used in \cite{Fodor:2001au}}
at Im$(\mu)=0$ in the phase transition point, i.e. at 
$\beta=5.04$ ($\approx 2500$ independent configurations) and at $\beta=5.085$ 
($\approx 7000$ independent configurations). From these two starting 
points with the use of the reweighting we tried to predict the plaquette 
($Pl$) and the $\bar{\psi}\psi$ expectation values and their uncertainties 
at $\beta=5.085$ and Im$(\mu)\neq 0$, that is at the target, imaginary 
chemical potential values.  We calculated the 
plaquette and the $\bar{\psi}\psi$ expectation values by 
(\ref{multi-parameter2}) at the target points, and we also used the new 
Metropolis-type method defined above which leads to  very similar results. 

The results are shown in Figure~\ref{fig:metro1} (see explanation there). 
The infinitely 
large errors of the 
single-parameter reweighting (Glasgow-method) indicate that the whole 
sample is thermalization, that is it does not provide information 
about the expectation value in the required point. 
In the right panel of Figure~\ref{fig:metro1} the 
second problem (lack of relevant configuration) mentioned above is seen. 

When we determine the EoS the jackknife errors are suitable to 
estimate the uncertainties of the reweighted quantities in an
appropriate region. We used the new error estimates only to provide the 
limit of the applicability of the reweighting procedure and of the 
jackknife method. 

\begin{figure}[t]
\centerline{\includegraphics[width=7cm]{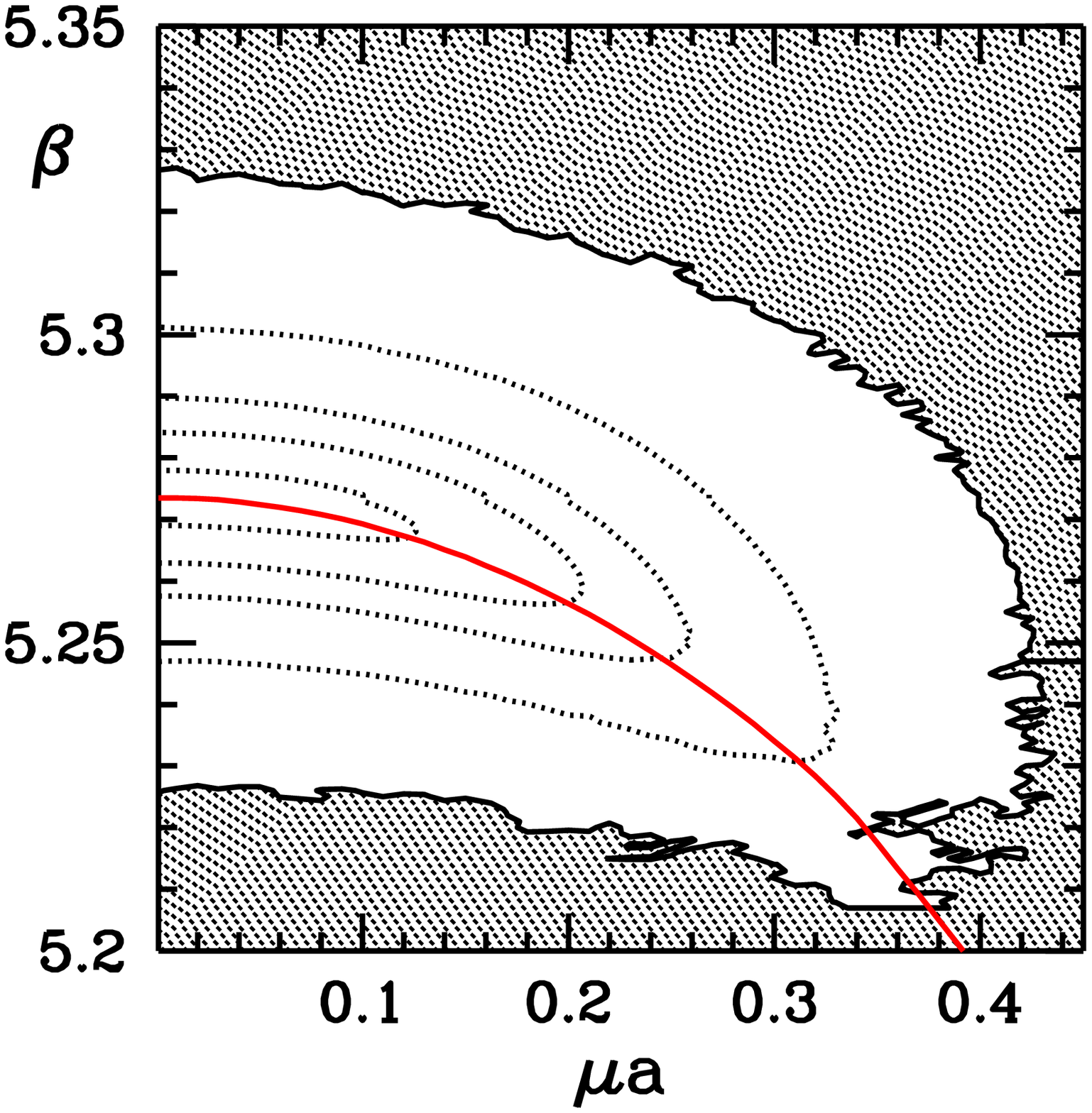} \hspace{5mm}
            \includegraphics[width=7cm]{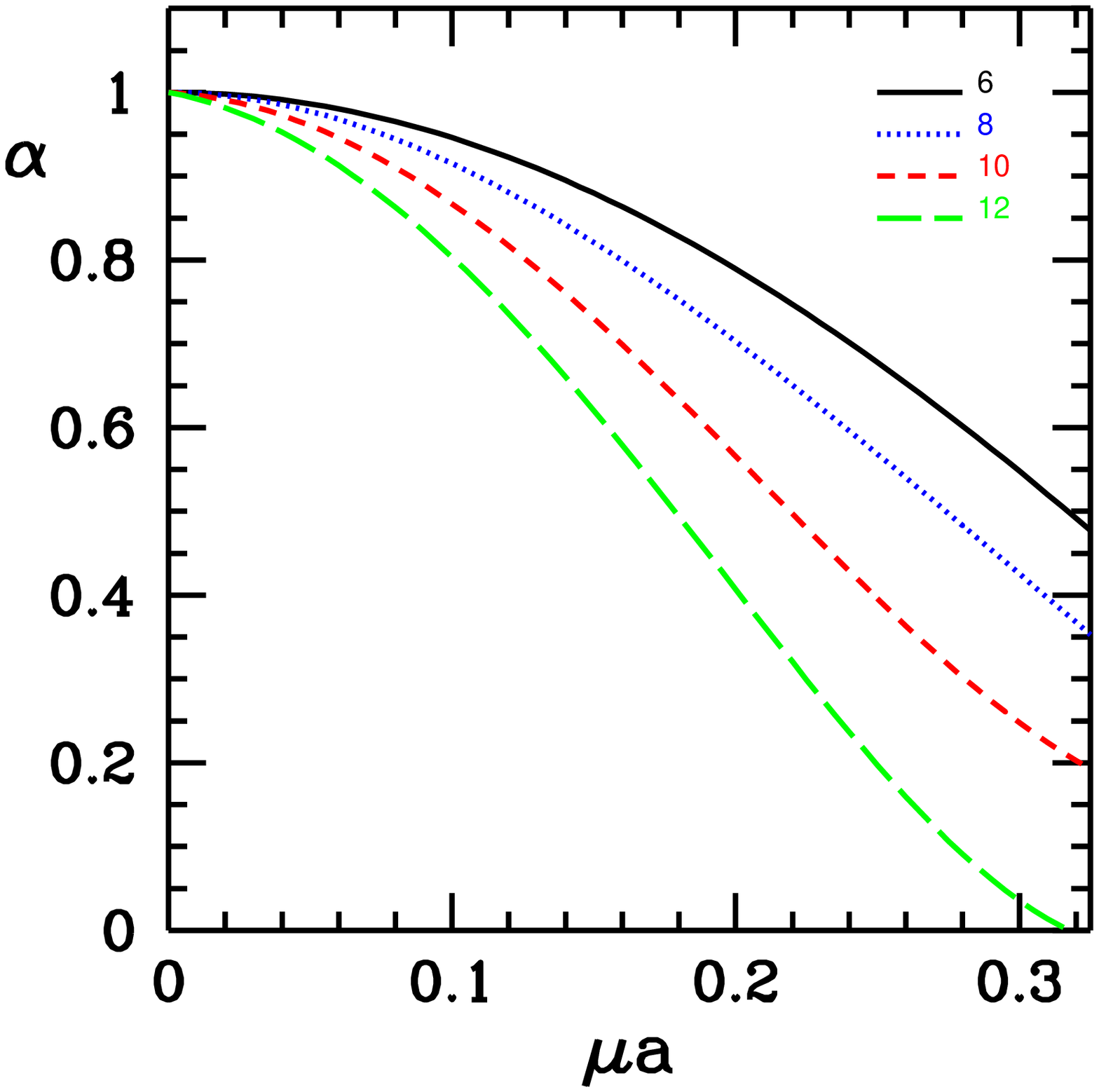}} 
\vspace{-2mm}
\caption{
(a) The left panel shows the real chemical potential--$\beta$ plane.  
33000 configurations were simulated at the parameter set at the 
critical $\beta$ in the $n_f=2+1$ flavour case. 
The dotted lines are the contours of the constant overlap. 
The dotted area is the unknown territory where the overlap vanishes.
The solid line is the phase transition line determined by the peaks  
of susceptibility. (b) In the right panel the volume and the $\mu$ 
dependence of the overlap ($\alpha$) is shown. Upper curves correspond 
to smaller lattice sizes, $4\cdot 6^3$, $4\cdot 8^3$, $4\cdot 10^3$ and 
$4\cdot 12^3$ respectively. 
\label{fig:contour}}
\end{figure}

We can define reweighting lines on the $\beta$--$\mu$ 
plane so as to make the least possible mistake during the reweighting 
procedure. To do this we introduce the notion of overlap measure which 
we denote by $\alpha$. The overlap measure is the normalised number of 
different configurations in the sample created with the  Metropolis-type 
reweighting after cutting off the thermalization. 
We plotted the contour lines of $\alpha$ 
in the left panel of Figure~\ref{fig:contour}. 
The dotted areas are unattainable, that means here the overlaps vanish, the 
errors are infinitely large. The best reweighting line can be defined for 
each simulation point. For a given value of $\mu$ we choose $\beta$ 
so that $\alpha$ be maximal. The points of the best reweighting lines are
given by the rightmost points of the contours of the overlap in 
Fig.~\ref{fig:contour}~(a).

Then the best reweighting 
lines are the contours of constant overlap or equivalently of constant error. 

The right panel of Figure~\ref{fig:contour} 
shows the $\mu$ dependence of the overlap at  fixed $\beta$ and  
quark mass parameters for different volumes ($V=4\cdot 6^3$, $4\cdot 
8^3$, $4\cdot 10^3$ and $4\cdot 12^3$). As expected, for fixed  $\mu$ 
larger volumes result in worse overlap. One can define the ``half-width'' 
($\mu_{1/2}$) of the $\mu$ dependence by the chemical potential value 
at which $\alpha=1/2$. One observes an approximate scaling behaviour 
for the half-width: $\mu_{1/2}\propto V^{-\gamma}$ with $\gamma \approx 1/3$. 

It is obvious that the two-parameter reweighting used  
previously does not follow  the LCP ($\beta$ gets smaller but 
the quark mass remains $m_0a$).   
Nevertheless, the best reweighting line along the LCP can be determined 
by two techniques.  One of them is the three-parameter reweighting, 
the other one is the interpolating method.  

As it can be seen in the left panel of Figure \ref{fig:contour}, the
change in $\beta$ is not very large for the two-parameter reweighting.
Therefore, one can remain on the LCP by a simultaneous, small change
of the mass parameter of the lattice action. This results in a
three-parameter reweighting (reweighting in $ma$, $\beta$ and
$\mu a$). Similarly to the two-parameter reweighting one can construct the 
best three-parameter reweighting line. 

Another possibility to stay on the line of constant physics at finite
$\mu$ is the interpolating technique.  One uses the two-parameter
reweighting for two LCP's and interpolates between them.  
The result of this method and the
predicition of the three-parameter reweighting agree quite well.  This
indicates that the requirement for the best overlap selects the same
weight lines even for rather different methods.

\section{Equation of state at non-vanishing chemical potential
\label{sec_muneq0}}

In this section we study the EoS at finite chemical potential. Since
we are interested in the physics of finite baryon density we use
$\mu=\mu_u=\mu_d\neq 0$ for the two light quarks and $\mu_s=0$ for the
strange quark. 

\begin{figure}[t]
\centerline{\includegraphics[width=8cm]{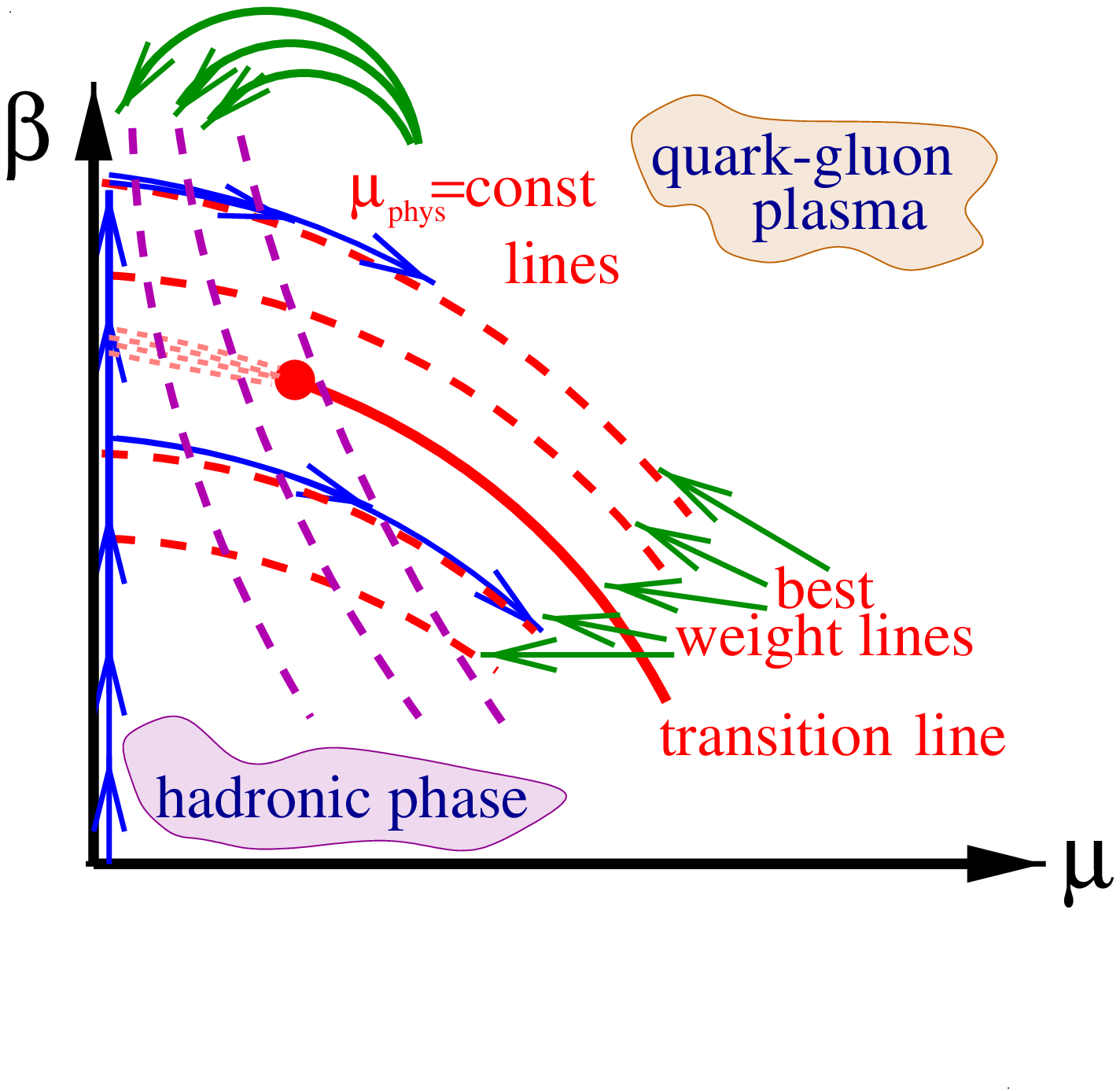}} 
\vspace{-2mm}
\caption{Illustration of the integral method at finite chemical
potential (left panel). The solid lines are $\mu={\rm const}$ lines on
the $\mu a - \beta$ plane. Dashed lines are the best reweighting lines
starting from different simulation points. Arrows show the path of
integration we used when evaluating eq.~(7.6).
\label{weightlines}}
\end{figure}

The pressure ($p$) can be obtained from the partition function
as $p$=$T\cdot\partial \log Z/ \partial V$ which can be written as
$p$=$(T/V) \cdot \log Z$ for large homogeneous systems.
On the lattice we can only determine the derivatives of $\log Z$ with respect
to the parameters of the action ($\beta, m, \mu$), so $p$ can be written as
a contour integral\cite{Engels:1990vr}:
\begin{eqnarray}
\frac{p}{T^4}&=&\frac{1}{T^3 V} \int d(\beta, m,\mu )
\left(
\left\langle \frac{\partial(\log Z)}{\partial \beta}\right\rangle,
\left\langle \frac{\partial(\log Z)}{\partial m}\right\rangle,
\left\langle \frac{\partial(\log Z)}{\partial \mu }\right\rangle\right).
\end{eqnarray}
The integral is by definition independent of the integration path.
The chosen integration paths are shown on Fig \ref{weightlines}.

The energy density can be written as
$\epsilon =(T^2/V)\cdot \partial(\log Z)/\partial {T}
+(\mu T/V)\cdot \partial(\log Z)/\partial\mu$.
By changing the lattice spacing $T$ and $V$ are simultaneously varied.
The special combination $\epsilon-3p$ contains only
derivatives with respect to $a$ and $\mu$:
\begin{equation}
\frac{\epsilon-3p}{T^4}=-\left.\frac{a}{T^3V}\frac{\partial \log(Z)}{\partial a}\right|_\mu
+\left. \frac{\mu}{T^3 V}\frac{\partial(\log Z)}{\partial\mu}\right|_a.
\end{equation}

\begin{figure}[t]
\centerline{\includegraphics[width=14cm]{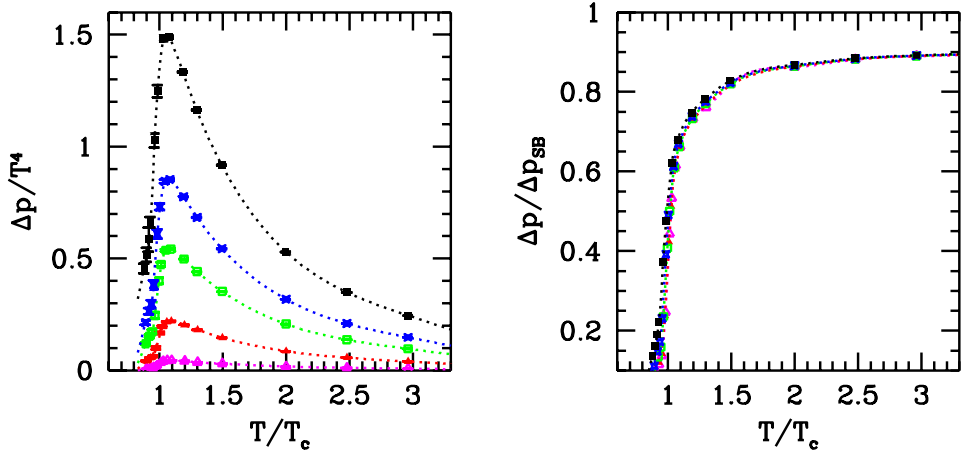}} 
\vspace{-2mm}
\caption{The equation of state at,
$\mu_B =100$, 210, 330, 410 and 530~MeV. The left panel shows
the pressure difference between the $\mu = 0$ and $\mu\neq 0$ cases
normalized by $T^4$, whereas on the right panel the normalization is
done by the $N_t=4$ lattice Stefan-Boltzman limit.
Note, that $\Delta p / \Delta p^{SB}$ seems to show
some scaling behaviour (it depends mostly on the temperature; whereas
its dependence on $\mu$ is much weaker). Thus, the $\mu$ dependence of
$\Delta p$ is almost completely given by the $\mu$ dependence of the
free gas.
\label{eosmu}}
\end{figure}

\begin{figure}[t]
\centerline{\includegraphics[width=14cm]{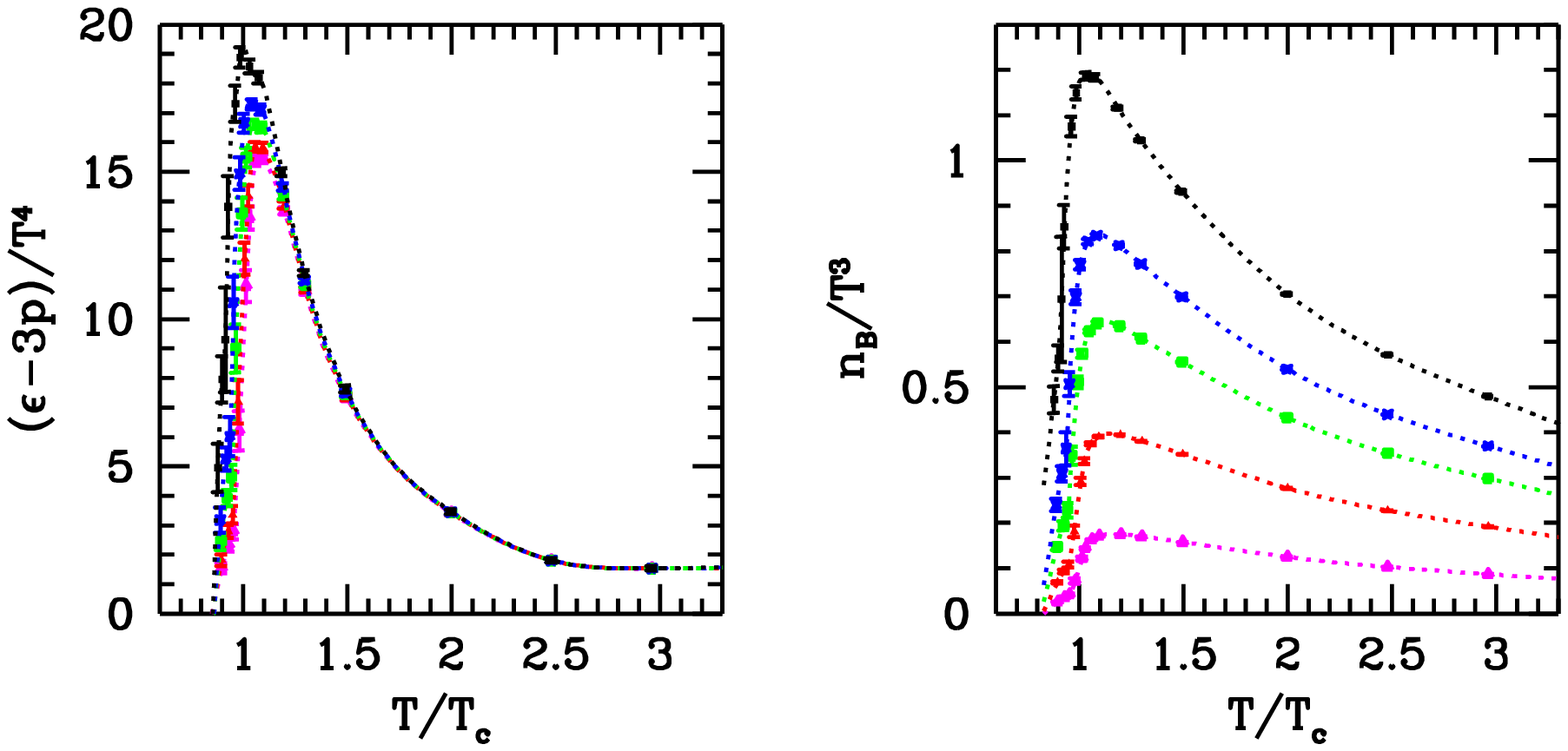}} 
\vspace{-2mm}
\caption{(a) $(\epsilon-3p)/T^4$ (b) Dimensionless baryon
number density as a function of $T/T_c$ at $\mu_B$=100,210,330,410~MeV and
$\mu_B$=530~MeV.
\label{eosmu3}}
\end{figure}

We present lattice results on $\Delta p(\mu,T)=p(\mu\neq
0,T)-p(\mu=0,T)$, $\epsilon(\mu,T)$-3$p(\mu,T)$ and $n_B(\mu,T)$. Our
statistical errorbars are also shown. They are rather small, in many
cases they are even smaller than the thickness of the lines.

On the left panel of Fig. \ref{eosmu} we present $\Delta p/T^4$ for
five different $\mu$ values. On the right panel normalisation is done
by $\Delta p^{SB}$, which is $\Delta p(\mu,T\rightarrow\infty)$.
Notice the interesting scaling behaviour.  $\Delta p / \Delta p^{SB}$
depends only on T and it is practically independent from $\mu$ in the
analysed region. The left panel of Fig. \ref{eosmu3} shows 
$\epsilon$-3$p$ normalised by $T^4$, which tends to zero for large
$T$. The right panel of Fig. \ref{eosmu3} gives the dimensionless
baryonic density as a function of $T/T_c$ for different $\mu$-s.

The  error coming from reweighting has been discussed previously.
Another source of error is the finiteness of the physical
volume. The volume dependence of physical observables
 is smaller than the statistical errors for the
plaquette average or  quark number density. 

\section{Conclusions, outlook}

We studied the thermodynamical properties of QCD at
finite chemical potential. We used the overlap improving
multi-parameter reweighting method. Our primary goal was to determine the
equation of state (EoS) on the line of constant physics (LCP) at
finite temperature and chemical potential.

We have pointed out that even at $\mu$=0 the EoS depends on the fact
whether we are on an LCP or not. According to our
findings pressure and $\epsilon$ - 3p (interacion measure) 
on the LCP have different high
temperature behaviour than in the ``non-LCP approach''. 

We discussed the reliability of the reweighting technique. We introduced 
an error estimate, which successfully shows the limits of the method 
yielding infinite errors in the parameter regions, where reweighting 
gives wrong results. We showed how to define and determine the best
weight lines on the $\mu$--$\beta$ plane.  

We discussed the two-parameter reweighting technique. Two techniques were
presented (three-parameter reweighting and the interpolating method) to 
stay on the LCP even when reweighting to non-vanishing chemical potentials.

We calculated the thermodynamic equations for $\mu \neq 0$ and
determined the EoS along an LCP. We presented lattice data on the 
pressure, the interaction-measure and the baryon number density as a function
of temperature and chemical potential. The physical range of our
analysis extended upto $500-600$~MeV in temperature and baryon chemical
potential as well.

Clearly much more work is needed to get the final form of
non-perturbative EoS of QCD. Extrapolation to the thermodynamic and continuum
limits is a very CPU demanding task in the $\mu \neq 0$ case. Physical
$m_\pi/m_\rho$ ratio should be reached by decreasing the light quark
mass. Finally, renormalised LCP's should be
used when evaluating thermodynamic quantites.

\section*{Acknowledgements}

This work was partially supported by Hungarian Scientific
grants, OTKA-T37615/\-T34980/\-T29803/\-M37071/\-OMFB1548/\-OMMU-708. 
For the simulations a modified version of the MILC
public code was used (see http://physics.indiana.edu/\~{ }sg/milc.html). 
The simulations were carried out on the 
E\"otv\"os Univ., Inst. Theor. Phys. 163 node parallel PC cluster.

\end{document}